\documentclass[aps,pra,twocolumn,showpacs,superscriptaddress]{revtex4-1}
\usepackage{graphicx}
\usepackage{bm}
\usepackage{amsmath,amsfonts,amssymb}
\usepackage[normalem]{ulem}
\usepackage{color}
\usepackage[usenames,dvipsnames,svgnames,table]{xcolor}

\newcommand{\mb}[1]{\mathbf{#1}}
\newcommand{\hatmb}[1]{\hat{\mathbf{#1}}}
\newcommand{\rr}{\mathbf{r}}
\newcommand{\kk}{\mathbf{k}}

\newcommand{\mt}[1]{\text{#1}}
\newcommand{\refeq}[1]{Eq.~(\ref{#1})}

\begin{document}

%\preprint{}

\title{Comprehensive analysis of the optical Kerr coefficient of graphene}


%\thanks{}
%\altaffiliation{Sandia National Laboratories, Livermore, CA 94550, USA}
\author{Daniel B. S. Soh}
%\email[]{dansoh@stanford.edu}
\affiliation{Edward L. Ginzton Laboratory, Stanford University, Stanford, CA 94305, USA}
\affiliation{Sandia National Laboratories, Livermore, CA 94550, USA}
\author{Ryan Hamerly}
\affiliation{Edward L. Ginzton Laboratory, Stanford University, Stanford, CA 94305, USA} 
\author{Hideo Mabuchi}
\affiliation{Edward L. Ginzton Laboratory, Stanford University, Stanford, CA 94305, USA} 


\date{\today}

\begin{abstract}
		We present a comprehensive analysis of the the nonlinear optical Kerr effect in graphene. We directly solve the S-matrix element to calculate the absorption rate, utilizing the Volkov-Keldysh-type crystal wave functions. We then convert to the nonlinear refractive index coefficients through the Kramers-Kronig relation. In this formalism, the source of Kerr nonlinearity is the interplay of optical fields that cooperatively drive the transition from valence to conduction band.  This formalism makes it possible to identify and compute the rates of distinct nonlinear processes that contribute to the Kerr nonlinear refractive index coefficient. The four identified mechanisms are two photon absorption, Raman transition, self coupling, and quadratic AC Stark effect. We also present a comparison of our theory with recent experimental and theoretical results. 
\end{abstract}

%\keywords{}

\maketitle

\section{Introduction}
The Kerr optical effect describes the change in the refractive index in a material due to the presence of a strong optical field.  It causes various nonlinear phenomena including self-focusing, self-phase modulation, and modulation instability. The Kerr effect can be used for optical bistability and all optical switching in the context of cavity nonlinear optics, which are the prominent building blocks for integrated and quantum photonics \cite{kirchmair2013observation, bermel2006single, lloyd1999quantum}.

Graphene's Dirac cones exhibit a linear dispersion, making it an excellent material for variety of electrical applications (see the review articles \cite{bonaccorso2010graphene, bao2012graphene}). More recently the ``giant'' optical nonlinearity of graphene has been explored by theory and experiments \cite{moss1990band, hendry2010coherent, zhang2012z, cheng2014third, wu2011purely, kumar2013third, hong2013optical, rioux2011current} (see a comprehensive review of recent progress in Cheng \it et al. \rm \cite{cheng2014third}). The third-harmonic generation and the Kerr-type intensity-dependent refractive index change have been investigated in those previous results. In general, near-band four-wave mixing and Kerr optical nonlinearity are similar processes while the third-harmonic generation is somewhat distinct process that involves waves that are widely separated in frequencies. Although there are a number of heuristic approaches including the Miller rule \cite{miller1964optical} to relate those two processes, the relation of the third-order susceptibilities describing the two processes are not rigorously proven, and this treatment generally assumes a direct-gap band structure with $\omega \ll E_g$ \cite{boyd2003nonlinear}. For graphene's optical nonlinearity, rigorous theories describing the third harmonic generation have been reported, including Mikhailov \cite{mikhailov2014quantum} who calculated the off-diagonal current density matrix for the interband transition and Cheng \it et al. \rm \cite{cheng2014third}, who used the semiconductor Bloch equation also to calculate the current density matrix. Although Cheng \it et al. \rm briefly showed the Kerr nonlinear coefficient result based on two-photon absorption, it is known that other mechanisms also contribute to the Kerr nonlinear coefficient \cite{sheik1991dispersion}. To understand the optical Kerr nonlinearity in graphene, it is thus necessary to exhaustively search for the physical mechanisms that contribute to the Kerr nonlinear coefficient. 

In this paper, we take the Sheik-Bahae approach that relies on the Kramers-Kronig relation to convert the absorption coefficients to dispersive coefficients, directly calculating the $S$-matrix element describing the transition from the valence to the conduction band, using the Volkov-Keldysh-type dressed-state wavefunctions \cite{sheik1991dispersion}. The advantage of this method, compared to others such as calculating the current density operators, is that it allows for complete identification of the contributing physical mechanisms through a closed-form analytic expansion that includes an arbitrary level of wave mixing processes. Using this method, we identified a full set of physical processes that include various separate two-photon transitions as well as energy band corrections, which also naturally appear in the formalism. For the first time that we are aware of, we successfully calculated the Kerr nonlinear coefficient of graphene as a function of optical frequency and produced an analytical formula. 

We first explain the strategy to calculate the Kerr nonlinear coefficient (section II). Once the formulation is laid out, we perform an exhaustive search for the two-photon transitions as well as energy band corrections (section III), producing a table of contributions. We then convert the contributions to finally calculate the Kerr nonlinear coefficients (section IV). A comparison between our theory and previous experimental and theoretical results is presented (section V), and conclusions follow (section VI).

\section{Calculation method}

Our method to calculate the Kerr nonlinear coefficient utilizes the Kramers-Kronig relation
\begin{equation}
	n(\omega) - 1 = \frac{c}{\pi} \text{P} \int_0^\infty \textbf{d} \omega' \frac{\alpha_r (\omega')}{\omega'^2 - \omega^2}, \label{eq:KK}
\end{equation}
where $n$ is the refractive index, $c$ is the speed of light, $\alpha_r$ is the absorption coefficient in unit (m$^{-1}$). The symbol P indicates that the integral is a Cauchy principal value. We note that the Kramers-Kronig theory is intrinsically a linear relation, allowing each contribution to be calculated separately and added up later. We follow Sheik-Bahae's argument \cite{sheik1991dispersion} to transform the above into a perturbative calculation relating high order terms between the absorption coefficient and the refractive index. Suppose that the absorption coefficient changes due to a presence of perturbing source at frequency $\Omega$. Then, the amount of change in absorption at frequency $\omega'$ can be written as $\Delta \alpha_r (\omega'; \Omega)$. \refeq{eq:KK} implies that the induced change in the real refractive index $\Delta n(\omega; \Omega)$ can be still similarly linearly related to the change $\Delta \alpha_r (\omega';\Omega)$:
\begin{equation}
	\Delta n (\omega; \Omega) = \frac{c}{\pi} \text{P} \int_0^\infty \text{d} \omega' \frac{\Delta \alpha_r(\omega'; \Omega)}{\omega'^2 - \omega^2}.
\end{equation}
We can also calculate the \it self-refractive index change, \rm setting $\Omega = \omega$:
\begin{equation}
\Delta n (\omega; \omega) = \Delta n(\omega) = \frac{c}{\pi} \text{P} \int_0^\infty \text{d} \omega' \frac{\Delta \alpha_r(\omega'; \omega)}{\omega'^2 - \omega^2}, \label{eq:KK-nonlinear}
\end{equation}
which describes the change of refractive index at frequency $\omega$, caused by the light of the same frequency. If one adopts a perturbative approach, this equation implies that the perturbative expansion of the refractive index is related to that of the absorption coefficient, order by order.

The Kerr nonlinear coefficient $n_2$ is defined as $\Delta n(\omega) = n_2 (\omega) I_\omega$ where $I_\omega$ is the intensity of frequency $\omega$. Hence, our task is to find an appropriate formula for $\Delta \alpha_r (\omega', \omega) \propto I_\omega$. Consider the amount of absorbed light at $\omega'$ from the absorption by the graphene. The light intensity difference due to the absorption is $\delta I_{\omega'} = - \alpha I_{\omega'}$, where $\alpha$ is the absorption rate. The absorption may change due to the presence of light at frequency $\omega$. The change in absorption is $\Delta \delta I_{\omega'} = - \Delta \alpha (\omega'; \omega) I_{\omega'}$. Here, $\Delta \alpha(\omega'; \omega)$ is the absorption rate change at $\omega'$, affected by the presence of light at $\omega$. The generated number of free-carriers from absorption is $N_f = - \mathcal{A} \cdot (\delta I_{\omega'}) \cdot \text{d} t / (\hbar \omega')$ where $\mathcal{A}$ is the area of graphene that is exposed to the light field, $\text{d} t$ is the time of exposure. The change of the number of free-carriers due to the presence of light at $\omega$ is $\text{d} N_f = - \mathcal{A} \cdot (\Delta \delta I_{\omega'}) \cdot \text{d} t / (\hbar \omega')$. Then, the change in the rate of free-carrier density due to the presence of light at $\omega$ is, using $\text{d} n_f = \text{d} N_f / \mathcal{A}$:
\begin{equation}
	\left( \frac{\text{d} n_f}{\text{d} t } \right) = \Delta \alpha(\omega'; \omega) \frac{I_{\omega'}}{\hbar \omega'}. \label{eq:absorption-transition-rate}
\end{equation}
We express such change perturbatively with respect to $I_\omega$, the intensity of light at $\omega$, and take the leading term. Then, we can express the changed amount of the free-carrier density generation rate by $(\text{d} n_f / \text{d} t) = A(\omega', \omega) I_{\omega} I_{\omega'}$ with an appropriate parameter $A(\omega', \omega)$. Then, we can find the effective absorption coefficient, dividing $\Delta \alpha (\omega'; \omega)$ by an effective atomic thickness $d_0$ of the graphene plane \cite{cheng2014third} to obtain
\begin{equation}
	\Delta \alpha_r (\omega'; \omega) = \frac{\Delta \alpha (\omega'; \omega)}{d_0} = \frac{\hbar \omega'}{d_0} A (\omega', \omega) I_{\omega}. \label{eq:absorption-coefficient}
\end{equation}
Finally this will allow us to obtain the Kerr coefficient $n_2$ using the relation in \refeq{eq:KK-nonlinear}:
\begin{equation}
	n_2 (\omega) = \frac{c}{\pi} \text{P} \int_0^\infty \text{d} \omega'  \left( \frac{\hbar \omega'}{d_0}\right)\frac{A(\omega', \omega)}{\omega'^2 - \omega^2}.
\end{equation}
Hence, to calculate $n_2$ it is now apparent that we need to calculate the change in the free-carrier density generation rate, expressed as a bilinear term $I_{\omega'} I_\omega$. 

The free-carrier density generation rate is related to the transition rate of the electron state between the valence band state and the conduction band state via the interaction Hamiltonian $H_{\mt{int}} = (e/m) \mathbf{p} \cdot \mathbf{A}$ with electron charge $e$, mass $m$, the canonical momentum $\mb{p}$, and the vector potential $\mb{A}$ of an external field:
\begin{align}
	&\left( \frac{\text{d} n_f}{\text{d} t }\right) = W \nonumber \\
	&= \frac{D}{\mathcal{A}} \sum_{\kk, \kk'}\Bigl( W_{v \rightarrow c} (\kk, \kk') f(v,\kk') (1 - f(c,\kk)) \nonumber \\
	& \qquad\qquad\quad -\ W_{c \rightarrow v} (\kk', \kk) (1 - f(v,\kk')) f(c, \kk) \Bigr),
\end{align}
where $W$ is the transition rate per area, $D = 4$ is the degeneracy factor of graphene (factor 2 from two disparate Bravais lattices and factor 2 from the two spin states), and $W_{i \rightarrow j} (\kk, \kk')$ is the transition rate from a Bloch wave with wavevector $\kk$ in the band $i$ to another with $\kk'$ in the band $j$. Here, $f(i, \kk)$ represents the Fermi distribution (occupation probability) of band $i$, for a given temperature and doping. If we restrict ourselves to the case of an ultrafast optical pulse so that we can ignore the thermal recombination rate, we can approximately set $W_{v \rightarrow c} (\kk, \kk') \simeq W_{c \rightarrow v} (\kk', \kk)$, which leads to
\begin{equation}
	\left( \frac{\text{d} n_f}{\text{d} t }\right) = \frac{D}{\mathcal{A}} \sum_{\kk, \kk'}W_{v \rightarrow c} (\kk, \kk') \left( f(v,\kk') - f(c, \kk)  \right). \label{eq:step1}
\end{equation}
The transition rate $W_{v \rightarrow c} (\kk, \kk')$ is calculated through the $S$-matrix formalism \cite{landau1977quantum}:
\begin{equation}
	W_{v \rightarrow c} (\kk, \kk') = \frac{\text{d}}{\text{d} t} |S_{cv} |^2, \label{eq:step2}
\end{equation}
where $\text{d}t$ is the incremental time of exposure, which is identical with the previous $\text{d} t$.  We use the first-order perturbation with respect to the interaction Hamiltonian and represent the $S$-matrix element:
\begin{eqnarray}
	S_{cv} &\simeq& - \frac{i}{\hbar} \int_{-t/2}^{t/2} \text{d} \tau \langle \psi_c (\kk') | \tilde{H}_{\text{int}} (\tau) | \psi_v (\kk) \rangle \nonumber \\
	&=& - \frac{i}{\hbar} \int_{-t/2}^{t/2} \text{d} \tau \langle \psi_c (\kk', \tau) | H_{\text{int}} (\tau) | \psi_v (\kk, \tau) \rangle. \label{eq:S_cv}
\end{eqnarray}
where $t$ is the total time of exposure, $|\psi_{c,v} (\kk) \rangle$ are the Dirac ket of the conduction and the valence band states with the Bloch wavevector $\kk$, and $\tilde{H}_{\text{int}} (\tau)$ is the interaction Hamiltonian in the \it interaction picture \rm from interaction Hamiltonian $H_{\text{int}} = (e/m) \mathbf{p} \cdot \mathbf{A} (\tau)$. 

A typical approach for calculating the third-order optical nonlinearity is to utilize the second order perturbation theory in the $S$-matrix formalism so that squaring it generates the third-order with respect to the interaction Hamiltonian. An alternative is to consider only the first-order perturbation and utilize the \it dressed state \rm wavefunction at time $\tau$ \cite{sheik1991dispersion}. For this, we need to obtain an expression for $\psi_{c,v} (\kk, \rr, \tau) = \langle \rr | \psi_{c,v} (\kk, \tau) \rangle$. 

Volkov first solved the Dirac equation for a free electron under the influence of an external field \cite{volkov1935class}, relying on the fact that the lack of binding potential makes the interaction Hamiltonian commute with the free unperturbed Hamiltonian $\mathbf{p}^2/2 m$. This allows for a relatively straightforward full solution. In a crystal lattice, however, one cannot use the same approach due to the presence of the binding potential of the lattice atoms. Keldysh in \cite{keldysh1965ionization} proposed a hybrid Volkov-type wave function for a Bloch wave inside a crystal lattice. Concretely, Keldysh's modified Volkov-type wavefunctions are
\begin{eqnarray}
	\psi_j (\kk, \rr, t) &=& u_j (\kk,\rr) \textbf{e}^{i \kk \cdot \rr} \nonumber \\
	&& \times  \exp \left[ - \frac{i}{\hbar} E_j(\mb{k}) t - \frac{i}{\hbar} \int_0^t \text{d} \tau E^\text{int}_j (\tau) \right], \nonumber \\
\end{eqnarray}
where $u_j (\kk, \rr)$ is the Bloch function that has the same period as the lattice, $j = c,v$ is the band index for either conduction or valance band respectively, and
\begin{eqnarray}
	E_c (\tau) &=& E_g + \frac{\hbar^2 \kk^2}{2 m_c},
	\quad E_v (\tau) = -\frac{\hbar^2 \kk^2}{2 m_v} , \nonumber \\
	E^{\text{int}}_j (\tau) &=& \frac{e \hbar }{m_j} \kk \cdot \mathbf{A} (\tau). \label{eq:quadratic-dispersion}
\end{eqnarray}
Keldysh's wavefunction is obtained by time-integrating the Schr\"{o}dinger equation after replacing the unperturbed Hamiltonian with energy eigenvalues (band energies). At the same time, one replaces the canonical momentum $\mathbf{p}$ with the crystal momentum $\hbar \mathbf{k}$, which are not identical in general, thereby, making the interaction Hamiltonian commute with it. Obviously if $t = 0$, the wavefunction reduces to a usual Bloch wavefunction. If the external field $\mathbf{A}$ is zero, the time evolution of the wavefunction follows the phase rotation based on the band energies. Here, replacing the mass with the effective masses $m_{c,v}$ is critical since the binding potential is effectively absorbed in the band-structure dependent effective mass as well as the band energies. As a result, the modified wavefunction appears as a free propagating wave with a modified energy through the first-order perturbation theory, periodically spatially modulated through the Bloch function $u_i (\kk, \rr)$. This is exactly what Bloch wavefunction is supposed to mean physically. 

This \it dressed state \rm formalism, as Sheik-Bahae \it et al. \rm pointed out, allows for a simple integral calculation of the $S$-matrix element, leading to identification of a complete set of physical mechanisms affecting the Kerr nonlinear coefficient, using only the first-order perturbation theory \cite{sheik1991dispersion}. This is the beauty of Keldysh's treatment of dressed state of the lattice crystal Bloch wavefunctions. The result was extremely successful that the experimental results from many semiconductors including both wide band-gaps and narrow band-gaps, spanning several orders of magnitude in $n_2$, match the calculated Kerr coefficient $n_2$ reasonably well (within a factor of 5). 

Graphene is radically different from the bulk semiconductors in multiple aspects, and thus, we cannot adopt Sheik-Bahae's approach directly. Beside the fact that graphene is a 2D material, Sheik-Bahae \it et al. \rm treated only the case for the typical quadratically dispersive materials with a non-zero bandgap energy, as shown in \refeq{eq:quadratic-dispersion}. This was the critical step, which allowed Sheik-Bahae \it et al. \rm to be able to calculate the Kramers-Kronig integration. Unfortunately graphene has a linear dispersion, the effective mass diverges, and the bandgap energy is zero.

To overcome the difficulty in treating graphene while retaining the spirit of Keldysh's crystal lattice full solution, we study whether we can replace the Hamiltonian usually appearing in the time integrated exponent with C-numbers. Let us investigate this possibility by setting
\begin{equation}
	|\psi_j (\kk,t) \rangle = \exp \left( - \frac{i}{\hbar} E_j (\kk) t - \frac{i}{\hbar} \int_{0}^t \text{d} \tau H_{\text{int}}  \right) |\psi_j (\kk) \rangle,
\end{equation} 
where $E_j (\kk)$ is the band energy and $|\psi_j (\kk) \rangle$ has a projection on to the position basis: the Bloch wavefunction
\begin{equation}
	\psi_i (\rr,\kk) = \langle \rr | \psi_i (\kk) \rangle = u_i (\rr, \kk) \text{e}^{i \kk \cdot \rr}.
\end{equation}
We put this into \refeq{eq:S_cv} and insert the closure relation
\begin{equation}
	\sum_{\kk''} |\psi_c (\kk'') \rangle \langle \psi_c (\kk'') | + |\psi_v (\kk'') \rangle \langle \psi_v (\kk'') | = \mathbf{1}.
\end{equation}
Noticing that $\sum_{m} \text{e}^{i (\kk - \kk'') \cdot \mathbf{R}_m} = N \delta_{\kk, \kk''}$ where $R_m$ is the $m$th lattice point and $N$ is the total number of atoms, and the $S$-matrix element is thus nonzero only when $\kk = \kk'$ (lattice momentum conservation), we obtain
\begin{eqnarray}
	S_{cv} =&&  - \delta_{\kk \kk'} \frac{i}{\hbar}\text{e}^{- i \omega_{cv}} \langle \psi_c(\kk) | \text{e}^{\frac{i}{\hbar} \int \text{d}\tau H_{\text{int}}} | \psi_c (\kk) \rangle \nonumber \\
	&& \times V_{cv}  \langle \psi_v(\kk) | \text{e}^{-\frac{i}{\hbar} \int \text{d}\tau H_{\text{int}}} | \psi_v (\kk) \rangle,
\end{eqnarray}
where
\begin{equation}
	V_{cv} = \langle \psi_c (\kk) | H_{\text{int}} | \psi_v (\kk) \rangle.
\end{equation}
To explain the above equation, we note that there are four terms appearing in $S_{cv}$ after putting the closure relation. We also note that $H_\text{int} \propto |\mb{A}|$ and all the terms in $S_{cv}$ are at least the first order in $H_\text{int}$. Since we are looking for the terms proportional to $I_\omega I_\omega'$ after squaring $S_{cv}$, and the intensity $I_\omega$ is proportional to $|\mb{A}|^2$, the terms we are interested in $S_{cv}$, before squaring, are at most first order in $H_\text{int}$. However, the terms involving $\langle \psi_c(\kk) | \text{e}^{\frac{i}{\hbar} \int \text{d} \tau H_{\text{int}}} | \psi_v(\kk) \rangle (= B)$ does produce only the second-order or higher in $H_\text{int}$, before squaring $S_{cv}$, because of the following reason. When perturbatively expanded, the leading orders of $B$ and $B^*$ are both the first order in $H_\text{int}$. Therefore, when multiplied by $H_\text{int}$ and other terms, the terms containing $B$ and $B^*$ result in the second order in $H_\text{int}$, before squaring $S_{cv}$. Again, since all terms are at least the first order in $H_\text{int}$, the terms containing $B$ and $B^*$ will result in the third order in $H_\text{int}$ or higher when squared. Therefore, the terms containing $B$ and $B^*$ are discarded in the above equation, and we are left with only one term.

This explicit $S$-matrix calculation implies that we may utilize the following wavefunction, which is correct in the first order perturbation theory of the interaction Hamiltonian: 
\begin{eqnarray}
	&&|\psi_j (\kk, t) \rangle = |\psi_j (\kk) \rangle \times \nonumber \\
	&&\exp \left( - \frac{i}{\hbar} E_j (\kk) t - \frac{i}{\hbar} \int_{0}^t \text{d} \tau \langle \psi_j (\kk) | H_{\text{int}} | \psi_j (\kk) \rangle  \right). \label{eq:volkov-graphene}
\end{eqnarray}
The remaining task is to evaluate the matrix elements of the interaction Hamiltonian in the basis $\{|\psi_{c} (\kk) \rangle, |\psi_v (\kk) \rangle \}$. 

Let us first note that
\begin{eqnarray}
	V_{ij} &=& \langle \psi_i (\kk) | \frac{e}{m} \mathbf{p} \cdot \mathbf{A} | \psi_j (\kk) \rangle \nonumber \\
	&=& \langle \psi_i (\kk) | \frac{e}{m} (m \mathbf{v} - e \mathbf{A}) \cdot \mathbf{A} | \psi_j (\kk) \rangle \nonumber \\
	&\simeq & e \mathbf{v}_{ij} \cdot \mathbf{A},
\end{eqnarray}
where $v_{ij}$ is the matrix element of the velocity operator
\begin{equation}
	v_{ij} = \langle \psi_i (\kk) | \mathbf{v} | \psi_j (\kk) \rangle.
\end{equation}
Here, we approximated the interaction Hamiltonian only up to the first order perturbation of the field $\mathbf{A}$, which is necessary to obtain the terms $I_\omega I_{\omega'}$. This is the critical aspect of our calculation method that makes the whole calculation procedure tractable.

The velocity operator, around the $\mathbf{K}$-points (Dirac cones) of the graphene, is obtained previously in \cite{cheng2014third}, where a sketch of derivation was presented. We present a concrete derivation in the appendix \ref{sec:velocity-matrix} since the elements of the velocity operator play an essential role in our calculation. The results around the Dirac cones are
\begin{equation}
	\mathbf{v}_{cc} = v_F \frac{\mathbf{q}}{|\mathbf{q}|},  \mathbf{v}_{vv} = - v_F \frac{\mathbf{q}}{|\mathbf{q}|}, \mathbf{v}_{cv} = - i v_F \frac{\hat{\mathbf{z}} \times \mathbf{q}}{|\mathbf{q}|}, \label{eq:velocity-matrix}
\end{equation}
with $\mathbf{v}_{vc} = \mathbf{v}_{cv}^*$. Here, $v_F$ is the Fermi velocity equivalent to $3 a t_{nn}/2 \hbar$ with $a = |\mathbf{a}|$ where $\mathbf{a}$ is the vector connecting the atom in the Bravais lattice $A$ to the nearest atom in the Bravais lattice $B$ ($a$ = 0.139 nm), and $t_{nn} \approx 2.7$ eV is the hopping energy between the nearest atom sites \cite{neto2009electronic}; $\hat{\mathbf{z}}$ is the direction of light propagation, perpendicular to the graphene 2D plane, and $\mathbf{q} = \mathbf{k} - \mathbf{K}$. We note that the velocity matrix is Hermitian.

Next, let us assume explicitly that the external field is represented as
\begin{equation}
\mathbf{A}(t) = \hat{\mathbf{a}} \sum_{i = 1}^{N_\omega} A_i \cos (\omega_i t + \phi_i),
\end{equation}
where $N_\omega$ is the number of light waves at different frequencies and $\hat{\mb{a}}$ is the unit vector such that $\hat{\mb{a}} \cdot \hat{\mb{z}} = 0$. Here, we assumed that all light fields at various wavelengths have the same polarization since we are primarily interested in the Kerr nonlinearity. The interaction Hamiltonian is, then,
\begin{equation}
	H_{\text{int}} = e \mathbf{v} \cdot \mathbf{A} = e \hat{\mathbf{a}} \cdot \mathbf{v} \sum_{i=1}^{N_\omega} A_i \cos (\omega_it + \phi_i).
\end{equation}
Then, the explicit wavefunction obtained by \refeq{eq:volkov-graphene} is
\begin{eqnarray}
	\psi_{c,v} &&(\kk,\rr, t) = u_{c,v} (\kk, \rr) \text{e}^{i \mb{k} \cdot \rr} \times \nonumber \\
	&&\text{exp} \left( - \frac{i}{\hbar} E_{c,v} (\mb{k}) t  + \sum_{j=1}^{N_\omega} i \eta_j^{c,v} \sin (\omega_j t + \phi_j) \right),
\end{eqnarray}
where
\begin{equation}
	\eta_j^{c,v} = - \frac{e A_j }{\hbar \omega_j} \hat{\mathbf{a}} \cdot \mathbf{v}_{cc,vv}.
\end{equation}
Finally we obtained an explicit form of the wave functions consisting in all C-numbers. This is one of the key steps in our procedure.

To evaluate the $S$-matrix element, we further use the Bessel function expansion:
\begin{equation}
	e^{i \eta \sin (\theta)} = \sum_{m=-\infty}^\infty J_m (\eta) \text{e}^{i n \theta}.
\end{equation}
After simplifying the time integral into Dirac delta functions, we finally obtain an explicit formula for the $S$-matrix element:
\begin{widetext}

	\begin{eqnarray}
		&&S_{cv} = - i \pi \delta_{\kk \kk'} e \hat{\mathbf{a}} \cdot \mathbf{v}_{cv} \sum_{i=1}^{N_\omega} A_i \sum_{n_1, \cdot, n_{N_\omega} = - \infty}^\infty \left( \prod_{j=1}^{N_\omega} J_{n_j} (\eta_j) \right) \left[ \begin{array}{l}\delta(-\hbar \omega_{cv} + \hbar \omega_i + \sum_{j=1}^{N_\omega} n_j \hbar \omega_j) \\
		~~+ \delta(-\hbar \omega_{cv} - \hbar \omega_i + \sum_{j=1}^{N_\omega} n_j \hbar \omega_j) \end{array} \right]. \label{eq:S-matrix-comprehensive}
	\end{eqnarray}
\end{widetext}
where
\begin{eqnarray}
	\eta_j &=& \frac{e A_{j}}{\hbar \omega_j} \hat{\mathbf{a}} \cdot (\mathbf{v}_{cc} - \mathbf{v}_{vv}), \nonumber \\
	\hbar \omega_{cv} &=&2 \hbar v_F |\mathbf{q}|.
\end{eqnarray}
This is the most general form of the $S$-matrix using the first-order perturbation theory based on the dressed state formalism. 	

Now that we have established our calculation procedures and obtained the elements necessary to carry them out, we want to emphasize that what we are searching for is the term proportional to $I_{\omega_2}$ in calculating the absorption coefficient for frequency $\omega_1$ (see \refeq{eq:absorption-coefficient}). Since the absorption coefficient is derived from the free-carrier transition rate, which is proportional to the square of the $S$-matrix element in \refeq{eq:S-matrix-comprehensive}, what we are searching for in the $S$-matrix elements are the terms that are proportional to $A_1 A_2$ where $A_i$ is the field amplitude at frequency $\omega_i$, since the amplitude of the field and the intensity are related as $I_{\omega_i} = (c \varepsilon_0 \omega_i^2/2) |A_i|^2$ where $\varepsilon_0$ is the vacuum permittivity. These terms eventually contribute to the Kerr nonlinear coefficient through the Kramers-Kronig relation in \refeq{eq:KK-nonlinear}. Therefore, we are primarily interested in searching for terms proportional to $A_1 A_2$ in the $S$-matrix.

In summary, we will first calculate the $S$-matrix element to obtain the transition rate from the valence to the conduction band via the interaction Hamiltonian, which is translated into the carrier density change rate and relates to the absorption coefficient. Then, we will evaluate the Cauchy principal value integration using the Kramers-Kronig relation, which finally leads to the Kerr nonlinear coefficient $n_2$. 

\section{Absorption coefficients}

In this section, we will calculate the absorption coefficients for both linear and nonlinear processes. We first verify that our formalism produces the well known result of the graphene linear absorption. Then, we proceed to the two-photon absorption and energy corrections, which are the essential stepping stones towards the Kerr nonlinear coefficient $n_2$.

\subsection{Linear absorption}

The linear absorption is the case where only one light wave is involved in the transition. Hence, we set $N_\omega =1$ and $n_1 = 0$ in the $S$-matrix element in \refeq{eq:S-matrix-comprehensive} to obtain
\begin{equation}
	S_{cv} = - i \pi \delta_{\kk \kk'} e \hat{\mathbf{a}} \cdot \mathbf{v}_{cv} A_1 \delta(- \hbar \omega_{cv} + \hbar \omega_1). \label{eq:linear-absorption-S}
\end{equation}
We note that the second delta function in  \refeq{eq:S-matrix-comprehensive} disappears since all frequencies are positive in our formalism. Using the typical treatment of square of the delta function \cite{cohen1992atom}:
\begin{equation}
	\delta^2 (\hbar \omega_{cv} - \hbar \omega_1) = \frac{T}{2 \pi \hbar} \delta (\hbar \omega_{cv} - \hbar \omega_1),
\end{equation}
we obtain the carrier density rate from the linear absorption using \refeq{eq:step1} and \refeq{eq:step2}
\begin{eqnarray}
	\left( \frac{\text{d} n_f}{\text{d} t}\right)_L &=& \frac{2 \pi e^2 A_1^2}{\hbar \mathcal{A}} \sum_{\kk} | \hat{\mathbf{a}} \cdot \mathbf{v}_{vc} |^2 \delta (\hbar \omega_{cv} - \hbar \omega_1) \nonumber \\
	&&\times (f(v,\kk) - f(c,\kk)).
\end{eqnarray}
We will replace the summation of $\kk$ into integral. For this, we need to evaluate the following using the previous result in \refeq{eq:velocity-matrix}:
\begin{equation}
	\langle | \hat{\mathbf{a}} \cdot \mathbf{v}_{cv}|^2 \rangle_{\kk} = \frac{1}{2 \pi} \int_0^{2 \pi} \text{d} \theta v_F^2 \cos^2 \theta = \frac{v_F^2}{2}.
\end{equation}
Converting the sum into integral ($(1/\mathcal{A}) \sum_{\kk} \rightarrow \int_\kk \text{d}^2 k / (2 \pi)^2$), and replacing $\hbar v_F q \rightarrow E$, we obtain
\begin{eqnarray}
	&&\left( \frac{\text{d} n_f}{\text{d} t} \right)_L \nonumber \\
	&=& \frac{\pi v_F^2 e^2 A_1^2}{\hbar} \int_0^\infty \frac{2 \pi q \text{d} q}{(2 \pi)^2} \delta(2 \hbar v_F q - \hbar \omega_1) (f(v,q) - f(c,q)) \nonumber \\
	&=& \frac{e^2 \omega_1}{8 \hbar^2} A_1^2 f_{vc} \left( \frac{\hbar \omega_1}{2}\right), \label{eq:delta-regulation}
\end{eqnarray}
with the occupation difference $f_{vc} (E) \equiv f \left(v,- E \right) - f \left(c, E \right)$. For an undoped graphene, we have $ f_{vc} (E) = \tanh ( E/2 k_B \mathcal{T})$ where $k_B$ is the Boltzmann constant and $\mathcal{T}$ is the temperature. Let us use the relation between the field amplitude $A_1$ and the intensity of the field $I_{\omega_1}$, given by $I_{\omega_1} = (c \varepsilon_0 \omega_1^2/2) A_1^2$. Using the relation in \refeq{eq:absorption-transition-rate} for the linear case, we finally obtain the linear absorption rate
\begin{equation}
	\alpha_0 = \frac{e^2}{4 \hbar c \varepsilon_0} \tanh \left( \frac{\hbar \omega_1}{4 k_B \mathcal{T}}\right). \label{eq:linear-absorption}
\end{equation}
This exactly matches the well-known broadband graphene linear absorption 2.3 \% in the optical frequencies ($\hbar \omega_1 \gg k_B \mathcal{T}$) \cite{kuzmenko2008universal,  bonaccorso2010graphene}.

It is noteworthy that the effect of doping, and hence, the non-zero Fermi energy $E_F$, is dramatically pronounced for low frequencies (long wavelengths), which exhibit a non-analytic behavior at low temperature. The treatment in \cite{wunsch2006dynamical} explicitly approximated the Fermi distribution functions at low temperatures and demonstrated such non-conventional behavior at longer wavelengths ($\hbar \omega_1 \ll 2 E_F$) in the low temperature limit. Such effect, however, is relatively small at optical frequencies.

\subsection{Two-photon absorption}

\begin{figure}[!tb]
	\centering
	\includegraphics[width=0.25\textwidth]{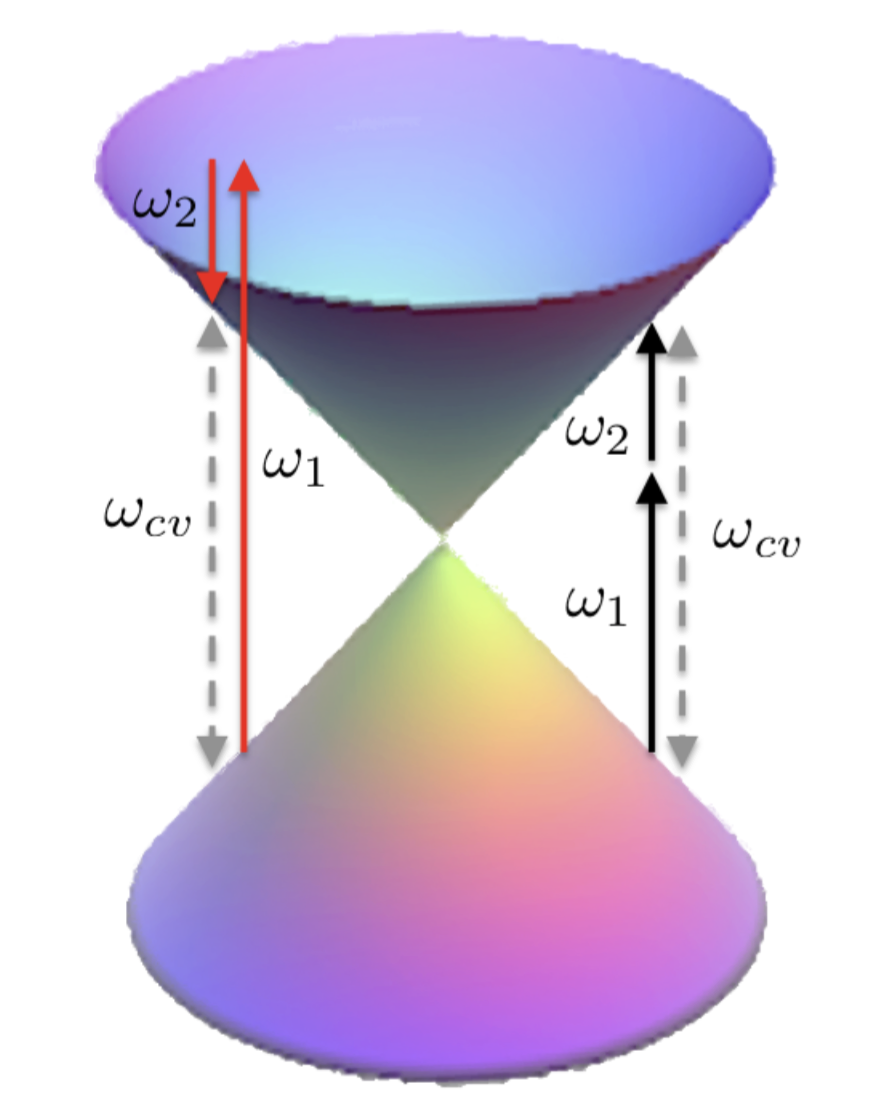}
	\caption{Two kinds of possible two-photon absorption transitions. Left: Raman transition, right: typical two-photon absorption.}
	\label{fig:transitions}
\end{figure}

The case corresponding to the two photon absorption is $N_\omega = 2$ (that is, two light frequencies are involved). If we use the Bessel function's Taylor series expansion to the lowest order:
\begin{equation}
	J_m (\eta) \simeq \frac{\eta^m}{2^m m!},
\end{equation}
we expect that, in the $S$-matrix element \refeq{eq:S-matrix-comprehensive}, $n_j$ corresponds to the number of times $A_j$ is multiplied. Recall that we are searching for the terms in the $S$-matrix element, which are proportional to $A_1 A_2$. These terms are obtained if in \refeq{eq:S-matrix-comprehensive}, for $i=1$, we set $n_1 = 0$ and $n_2 = 1$ while, for $i=2$, we set $n_1 = 1$ and $n_2 = 0$. The delta functions enforce that only three cases are possible, satisfying:
\begin{eqnarray}
	- \hbar \omega_{cv} + \hbar\omega_1 + \hbar \omega_2 &=& 0, \nonumber \\
	- \hbar \omega_{cv} - \hbar\omega_1 + \hbar \omega_2 &=& 0, \nonumber \\
	- \hbar \omega_{cv} + \hbar\omega_1 - \hbar \omega_2 &=& 0. \label{eq:two-photon-mechanisms}
\end{eqnarray}
Figure \ref{fig:transitions} shows the transitions that the above three cases represent. The first transition describes the case where $\omega_1, \omega_2 < \omega_{cv}$, which is the typical two-photon absorption. The second and the third transitions describe the Raman transitions where one frequency is larger than the band energy difference. 

We will first calculate the absorption coefficient for the case where both frequencies are smaller than the band transition energy. The $S$-matrix element is
\begin{eqnarray}
	S_{cv} &=& - i \delta_{\kk \kk'} \frac{\pi e^2}{2} (\hat{\mathbf{a}} \cdot \mathbf{v}_{cv}) (\hat{\mathbf{a}} \cdot (\mathbf{v}_{cc} - \mathbf{v}_{vv})) \left( \frac{1}{\hbar \omega_1} + \frac{1}{\hbar \omega_2}\right) \nonumber \\
	&& \times A_1 A_2 \delta (- \hbar \omega_{cv} + \hbar \omega_1 + \hbar \omega_2).
\end{eqnarray}
To proceed further to calculate the change of the free-carrier density generation rate, we must evaluate the following, using \refeq{eq:velocity-matrix},
\begin{eqnarray}
	&&\langle | (\hat{\mathbf{a}} \cdot \mathbf{v}_{cv}) (\hat{\mathbf{a}} \cdot (\mathbf{v}_{cc} - \mathbf{v}_{vv}))|^2 \rangle_\kk\nonumber \\
	&=& \frac{1}{2 \pi} \int_0^{2\pi} \text{d} \theta ~ 4 v_F^2 | \cos \theta \sin \theta |^2 = \frac{v_F^2}{2},
\end{eqnarray}
where we considered the fact that $\mathbf{v}_{ii}$ is orthogonal to $\mathbf{v}_{cv}$ while they both reside in the graphene 2D plane. Using this and after a long but elementary calculation, we obtain
\begin{equation}
	\left(  \frac{\text{d} n_f}{\text{d} t}\right)_{\text{TPA}} = \frac{e^4 v_F^2}{32 \hbar^3} \frac{(\hbar \omega_1 + \hbar \omega_2)^3}{(\hbar \omega_1)^2 (\hbar \omega_2)^2} A_1^2 A_2^2 f_{vc} \left( \frac{\hbar (\omega_1 + \omega_2)}{2} \right).
\end{equation}
Converting the field amplitudes into the intensities and using \refeq{eq:absorption-transition-rate}, we obtain the absorption coefficient
\begin{eqnarray}
	\Delta \alpha_r^{\text{TPA}} (\omega_1; \omega_2) &=& \frac{9 a^2 e^4}{32 c^2 d_0 \hbar \varepsilon_0^2 t_{nn}^2} \frac{(\hbar \omega_1/t_{nn} + \hbar \omega_2/t_{nn})^3}{(\hbar \omega_1/t_{nn})^3 (\hbar \omega_2/t_{nn})^4} \nonumber \\
	&&\times f_{vc} \left( \frac{\hbar (\omega_1 + \omega_2)}{2} \right) I_{\omega_2}.
\end{eqnarray}

The degenerate case where $\omega_1 = \omega_2$ is of particular interest. It immediately follows from the above equation that the degenerate two-photon absorption rate (which is the absorption coefficient multiplied by $d_0$) is
\begin{equation}
	\Delta \alpha^{\text{DTPA}} (\omega) = \frac{9 a^2 e^4}{4 c^2 \hbar \varepsilon_0^2 t_{nn}^2} \left( \frac{t_{nn}}{\hbar \omega}\right)^4 f_{vc} \left( \hbar \omega\right) I_\omega. \label{eq:101}
\end{equation}
Note that the two photon absorption is proportional to the inverse fourth power of the optical frequency $\omega$. This result agrees precisely with the reference \cite{cheng2014third}. 

\subsection{Raman transition absorption}

As we indicated in the previous section, another cooperative two-photon absorption is possible through the second, and the third case in \refeq{eq:two-photon-mechanisms}. For the case where $\omega_2 > \omega_{cv}$ (the second equation in \refeq{eq:two-photon-mechanisms}), we obtain a term in the $S$-matrix that is proportional to $A_1 A_2$ by setting $N_\omega = 2$ and $n_1 = 0, n_2 = 1$. We ignore the case where either $n_1$ or $n_2$ becomes negative since those terms do not produce the terms proportional to $A_1 A_2$. Also for the other case where $\omega_1 > \omega_{cv}$, we have the $S$-matrix element proportional to $A_1 A_2$ by setting $N_\omega = 2$ and $n_1 = 1, n_2 = 0$. Then, for these two cases, we obtain
\begin{eqnarray}
	S_{cv}^{\text{RT,i}} &=& i \delta_{\kk \kk'} \frac{\pi e^2}{2 \hbar \omega_{i}} (\hat{\mathbf{a}} \cdot \mathbf{v}_{cv}) (\hat{\mathbf{a}} \cdot (\mathbf{v}_{cc} - \mathbf{v}_{vv})) \nonumber \\
	&& \times A_1 A_2 \delta (- \hbar \omega_{cv} - \hbar \omega_j + \hbar \omega_i),
\end{eqnarray}
where $i,j = 1,2$ with $i \neq j$. We need to be careful in calculating the absorption coefficient since we will be using \refeq{eq:absorption-transition-rate} where $I_{\omega_1} / \hbar \omega_1$ appears on the right hand side. The result is that
\begin{eqnarray}
	\Delta \alpha_r^{\text{RT,1}} (\omega_1, \omega_2) &=& \frac{9 a^2 e^4}{32 c^2 d_0 \varepsilon_0^2 \hbar t_{nn}^2} \left(\frac{(\hbar \omega_2 / t_{nn}) - (\hbar \omega_1 / t_{nn})}{(\hbar \omega_1/t_{nn}) (\hbar \omega_2 / t_{nn})^4}\right) \nonumber \\
	&& \times f_{vc} \left( \frac{\hbar (\omega_2 - \omega_1)}{2 }\right) I_{\omega_2}, \nonumber \\
	\Delta \alpha_r^{\text{RT,2}} (\omega_1, \omega_2) &=& \frac{9 a^2 e^4}{32 c^2 d_0 \varepsilon_0^2 \hbar t_{nn}^2} \left(\frac{(\hbar \omega_1 / t_{nn}) - (\hbar \omega_2 / t_{nn})}{(\hbar \omega_1/t_{nn})^3 (\hbar \omega_2 / t_{nn})^2}\right) \nonumber \\
	&& \times f_{vc} \left( \frac{\hbar (\omega_1 - \omega_2)}{2}\right) I_{\omega_2}.
\end{eqnarray}
We note that $\omega_2 > \omega_{cv} > \omega_1$ corresponds to $\Delta \alpha_r^{\text{RT,1}}$ whereas $\omega_1 > \omega_{cv}> \omega_2$ corresponds to $\Delta \alpha_r^{\text{RT,2}}$. Both absorption coefficients are thus always positive.

\subsection{Self-coupling}

Another term in $S$-matrix element that is proportional to $A_1 A_2$ is through expanding the zeroth order Bessel function into
\begin{equation}
	J_0 (\eta) \simeq 1 - \frac{\eta^2}{4},
\end{equation}
in the linear absorption calculation in \refeq{eq:linear-absorption-S}. That is, when calculating the linear absorption at $\omega_1$, the presence of light of frequency $\omega_2$ may affect the absorption through the second-order expansion term in the above equation. Physically the process can be understood as a reduction in the oscillation strength at $\omega_1$ by renormalizing the interband coupling due to the acceleration of the free-carriers caused by the field at $\omega_2$ \cite{sheik1991dispersion}. The result is the modified $S$-matrix element from the linear absorption \refeq{eq:linear-absorption-S}
\begin{equation}
	S_{cv} = - i \pi e \hatmb{a} \cdot \mb{v}_{cv} A_1 \left(1 - \frac{\eta_2^2}{4} \right) \delta ( - \hbar \omega_{cv} + \hbar \omega_1),
\end{equation}
where $\eta_2 = e A_2 \hatmb{a} \cdot (\mb{v}_{cc} - \mb{v}_{vv}) / \hbar \omega_2$. Squaring it produces the term that is proportional to $A_1^2 A_2^2$, while we discard the higher order term proportional to $A_1^2 A_2^4$. Through the similar calculation procedure as other mechanisms, we obtain the absorption coefficient from the self-coupling
\begin{equation}
	\Delta \alpha_r^{\mt{SC}} (\omega_1 ; \omega_2) = - \frac{9 a^2 e^4}{8 c^2 d_0 \varepsilon_0^2 \hbar t_{nn}^2} \left( \frac{t_{nn}}{\hbar \omega_2} \right)^4 f_{vc} \left( \frac{\hbar \omega_1}{2}\right) I_{\omega_2}.
\end{equation}
As expected, the self-coupling results in the reduction of the absorption (the quantity above is negative). 

\subsection{Quadratic AC Stark effect}
The last remaining contribution that produces a term proportional to $I_{\omega_1} I_{\omega_2}$ can be traced by recalling \refeq{eq:delta-regulation} where the argument inside the delta function plays a role. That is, integrating $E \; \delta (2 E - 2 E')$ with respect to $E = \hbar v_F k$ results in a factor $E'$. This implies that, if for some reason the argument of the delta function is modified to $2E - 2 E' + \Delta E'$, the factor appearing after integration must be $(E' - \Delta E'/2)$, in the change of the free-carrier density generation rate as well as in the absorption coefficient. Particularly if such modification involves a term that is proportional to $A_2^2$ of light at frequency $\omega_2$, a new term that is proportional to $A_1^2 A_2^2$ appears, which must be counted for to calculate the Kerr nonlinear coefficient. 

The AC Stark effect is such a quadratic energy modification for both the conduction and the valence band. The energy shift adds more on the band energy difference by \cite{autler1955stark}
\begin{equation}
	\Delta E' = \frac{\hbar \Omega^2}{2 \Delta} = -\frac{e^2}{2(\hbar \omega_1 - \hbar \omega_2)} |\hatmb{a} \cdot \mb{v}_{cv}|^2 A_2^2,
\end{equation}
where $\Omega$ is the Rabi oscillation frequency and $\Delta$ is the detuning of $\omega_2$ light from the main resonance at $\omega_1$. Note that the band energy may increase or decrease,  depending on whether one has blue or red detuning. This quadratic AC Stark shift produces a contribution that is proportional to $A_1^2 A_2^2$ in the change of the free-carrier density generation rate. The result appearing in the absorption coefficient is, after integral calculations,
\begin{eqnarray}
	&&\Delta \alpha_r^{\mt{QSE}} (\omega_1; \omega_2) =\nonumber \\
	&& - \frac{9 a^2 e^4}{16 c^2 d_0 \hbar \varepsilon_0^2 t_{nn}^2} \left( \frac{t_{nn}}{\hbar \omega_2}\right)^2 \left( \frac{t_{nn}}{\hbar \omega_1}\right) \nonumber \\
	&& \times \left( \frac{1}{(\hbar \omega_1/t_{nn}) - (\hbar \omega_2 / t_{nn})} \right) f_{vc} \left( \frac{\hbar \omega_1}{2}\right) I_{\omega_2}. \label{eq:AC-stark}
\end{eqnarray}
Note that the overall sign depends on the sign of $\omega_1 - \omega_2$. Hence, the quadratic AC Stark effect increases (decreases) the absorption rate if $\omega_1 > \omega_2$ ($\omega_2 > \omega_1$), respectively.

\subsection{Summary of absorption coefficients}

Let us define a lumped parameter
\begin{equation}
	\alpha_2 = \frac{9 a^2 e^4}{32 c^2 d_0 \varepsilon_0^2 \hbar t_{nn}^2} = 7.772 \times 10^{-11} ~\mt{(m/W)},
\end{equation}
where we used $d_0 = 0.33$ nm \cite{cheng2014third}. We can express the various absorption coefficients using the following convention:
\begin{equation}
	\Delta \alpha_r (\omega_1; \omega_2) = \alpha_2 F \left( \frac{\hbar \omega_1}{t_{nn}}; \frac{\hbar \omega_2}{t_{nn}}\right) \tilde{f}_{vc} \left( \frac{\hbar \omega_1}{2}, \frac{\hbar \omega_2}{2} \right) I_{\omega_2}. \label{eq:generic-absorption}
\end{equation}

\begin{table}%[H] add [H] placement to break table across pages
	\caption{Summary of absorption coefficient function $F$ and occupation function $\tilde{f}_{vc}$ used in \refeq{eq:generic-absorption} for various two-photon absorption mechanisms.}\label{tab:absorption}
	\begin{ruledtabular}
	\begin{tabular}{ccc}
		Contribution & $F(x_1; x_2)$ & $\tilde{f}_{vc} (y_1,y_2)$ \\*[2mm]
		\hline
		\\*[-3mm]
		Two photon absorption & $\cfrac{(x_1 + x_2)^3}{x_1^3 x_2^4}$ & $f_{vc} \left( y_1 + y_2\right)$ \\*[3mm]
		Raman transition ($\omega_2 > \omega_1$) & $\cfrac{x_2 - x_1}{x_1 x_2^4}$ & $f_{vc} \left( y_2 - y_1 \right)$ \\*[3mm]
		Raman transition ($\omega_1 > \omega_2$) & $\cfrac{x_1 - x_2}{x_1^3 x_2^2}$ & $f_{vc} (y_1 - y_2)$ \\*[3mm]
		Self-coupling & $- \cfrac{4}{x_2^4}$ & $f_{vc} (y_1)$ \\*[3mm]
		Quadratic AC Stark shift & $-\cfrac{2}{x_1 x_2^2 (x_1 - x_2)}$ & $f_{vc} (y_1)$
% Lines of table here ending with \\
	\end{tabular}
	\end{ruledtabular}
\end{table}

Table \ref{tab:absorption} summarizes our calculated absorption coefficients from the various two-photon mechanisms. It is worth noting that the contribution from the self-coupling and the quadratic AC Stark shifts are the corrections to the linear absorption calculated in \refeq{eq:linear-absorption}, and thus the negative sign does not mean creation of photons. Also, the occupation probability $\tilde{f}_{vc}$ reflects the energy difference of the actual conduction and the valence bands involved in each process. 

More practically, we note that the nonlinear sheet absorption $\Delta \alpha (\omega_1; \omega_2) = d_0 \Delta \alpha_r (\omega_1; \omega_2)$ is actually a measurable quantity, and is obtained by
\begin{equation}
	\Delta \alpha (\omega_1 ; \omega_2) = \alpha_2' F \left( \frac{\hbar \omega_1}{t_{nn}}; \frac{\hbar \omega_2}{t_{nn}}\right) \tilde{f}_{vc} \left( \frac{\hbar \omega_1}{2}, \frac{\hbar \omega_2}{2} \right) I_{\omega_2},
\end{equation}
where $\alpha_2' = 2.565 \times 10^{-20}$ (m$^2$/W), which makes $\Delta \alpha (\omega_1; \omega_2)$ unitless.

\section{Kerr nonlinear coefficients}

We are now ready to calculate the Kerr nonlinear coefficients $n_2$ using the absorption coefficients in \refeq{eq:generic-absorption} with the TABLE \ref{tab:absorption}, through evaluating the Kramers-Kronig relation in \refeq{eq:KK-nonlinear}. We need to carry out the Cauchy principal value integration with respect to $\omega_1$. Using the relation $\Delta n(\omega_2) = n_2 I_{\omega_2}$ in \refeq{eq:KK-nonlinear} and using \refeq{eq:generic-absorption}, we obtain, replacing $\hbar \omega_1/t_{nn} \rightarrow x_1$ and $\hbar \omega_2 / t_{nn} \rightarrow x_2$,
\begin{align}
	&n_2 (\omega_2) \nonumber \\
	&= \frac{c \alpha_2}{\pi} \mt{P} \int_0^\infty \frac{\mt{d} \omega_1}{\omega_1^2 - \omega_2^2} F \left( \frac{\hbar \omega_1}{t_{nn}} ; \frac{\hbar \omega_2}{t_{nn}} \right) \tilde{f}_{vc} \left( \frac{\hbar \omega_1}{2}, \frac{\hbar \omega_2}{2} \right) \nonumber \\
	&= \frac{c \hbar \alpha_2}{\pi t_{nn}} \mt{P} \int_0^\infty \frac{\mt{d} x_1}{x_1^2 - x_2^2} F(x_1; x_2) \tilde{f}_{vc} \left( \frac{t_{nn} x_1}{2}, \frac{t_{nn} x_2}{2}\right). \nonumber \\
\end{align}

We first calculate the contribution from the two-photon absorption. The presence of the occupation function $\tilde{f}_{vc}$ complicates the calculation. In principle, one can evaluate using a full numerical integration. However, with a reasonable assumption, we can carry out the integration analytically. For undoped graphene, the occupation function is $f_{vc} (E) = \tanh (E/2 k_B \mathcal{T})$. Particularly when $\omega_{1,2}$ are optical frequencies where $\hbar \omega_{1,2} \gg k_B \mathcal{T}$ with a reasonable temperature range well below the Fermi temperature, one can safely approximate $f_{vc} (\hbar (\omega_1 + \omega_2)/2) \simeq 1$. Then we obtain
\begin{equation}
	n_2^{\mt{TPA}}(\omega_2) = \frac{c \hbar \alpha_2}{\pi t_{nn}} \left( \frac{t_{nn}}{\hbar \omega_2}\right)^4 \mt{P} \int_0^\infty \mt{d} x_1 \frac{(x_1 + x_2)^3}{x_1^3 (x_1^2 - x_2^2)}. \label{eq:n2-TPA-step1}
\end{equation}

One can express the Kramers-Kronig relation between the real and the imaginary values of the susceptibility $\chi = \chi' + i \chi''$:
\begin{equation}
	\chi' (\omega) = \mt{P} \int_{-\infty}^\infty \frac{\mt{d} \nu}{\pi} \frac{\chi'' (\nu)}{\nu - \omega} = \mt{P} \int_0^\infty \frac{\mt{d} \nu}{\pi} \frac{2 \nu \chi'' (\nu)}{\nu^2 - \omega^2},
\end{equation}
where the real part $\chi'$ is related to $\Delta n (\omega)$ and the imaginary part $\chi''$ is related to $\Delta \alpha_r (\nu; \omega)$.  This is the same expression as \refeq{eq:KK} and \refeq{eq:KK-nonlinear}. Due to the condition that $\chi''(-\omega) = - \chi'' (\omega)$ that comes from the realness of the susceptibility $\chi(t)$ (i.e., $\chi^* (\omega) = \chi(-\omega)$), any singularity occuring around $\nu=0$ is canceled due to the anti-symmetry between $\nu=0^+$ and $\nu = 0^-$ of the integrand. Hence, on these physical grounds we can safely discard any mathematical singularity at $x_1 = 0$ in the evaluation of the Cauchy principal integral \refeq{eq:n2-TPA-step1}. Integrating with respect to $x_1$ in the separate intervals $[x_0, x_2-\epsilon]$ and $[x_2 + \epsilon, \infty)$, taking the limit $x_0 \rightarrow 0$ and then $\epsilon \rightarrow 0$, and discarding the singularity occurring at $x_1=0$, we calculate
\begin{equation}
	\mt{P} \int_0^\infty \mt{d} x_1 \frac{(x_1 + x_2)^3}{x_1^3 (x_1^2 - x_2^2)} = - 4 \frac{\ln x_2}{x_2}.
\end{equation}
From this we obtain the contribution of the two-photon absorption to the Kerr nonlinear coefficient as
\begin{equation}
	n_2^{\mt{TPA}} (\omega_2) = \frac{4 c \hbar \alpha_2}{\pi t_{nn}} \left(\frac{t_{nn}}{\hbar \omega_2}\right)^5 \ln \left( \frac{t_{nn}}{\hbar \omega_2}\right).
\end{equation}
We note that this quantity is positive if $ \hbar \omega_2 < t_{nn} $, but negative otherwise. We also note that the unit of this quantity is (m$^2$/W) such that $n_2 (\omega_2) I_{\omega_2}$ becomes unitless.

Calculating the Kerr coefficient from the two Raman transition processes requires to evaluate the following:
\begin{eqnarray}
	n_2^{\mt{RT}} (x_2) &=& \frac{c \hbar \alpha_2}{\pi t_{nn}} \times \nonumber \\
	&& \left[ \mt{P} \int_0^{x_2} \mt{d} x_1 \frac{x_2 - x_1}{x_1 x_2^4 (x_1^2 - x_2^2)} f_{vc} (t_{nn} x_2 - t_{nn} x_1) \right. \nonumber \\
	&& \left. + \mt{P} \int_{x_2}^\infty \mt{d} x_1 \frac{x_1 - x_2}{x_1^3 x_2^2 (x_1^2 - x_2^2)} f_{vc} (t_{nn} x_1 - t_{nn} x_2) \right]. \nonumber \\
\end{eqnarray}
Since we are concerned about optical frequencies $\omega_1, \omega_2$, it is also a reasonable assumptions to set
\begin{equation}
	f_{vc} (x) \simeq \theta(x) = \left\{ \begin{array}{ll} 1, & x \ge 0, \\ 0, & x<0. \end{array} \ \right.
\end{equation}
where $\theta (x)$ is the Heaviside step function. Then, the above reduces to
\begin{eqnarray}
	n_2^{\mt{RT}} (x_2) \simeq \frac{c \hbar \alpha_2}{\pi t_{nn}}  && \left[  \mt{P} \int_0^{x_2} \mt{d} x_1 \frac{x_2 - x_1}{x_1 x_2^4 (x_1^2 - x_2^2)} \right. \nonumber \\
	&& ~\left. + \mt{P} \int_{x_2}^\infty \mt{d} x_1 \frac{x_1 - x_2}{x_1^3 x_2^2(x_1^2 - x_2^2)} \right].
\end{eqnarray}
Ignoring the singularity occurring at $x_1 = 0$, we obtain
\begin{eqnarray}
	\mt{P} \int_0^{x_2} \mt{d} x_1 \frac{x_2 - x_1}{x_1 x_2^4 (x_1^2 - x_2^2)} &=&  - \frac{\ln x_2 }{x_2^5} + \frac{\ln 2}{x_2^5}, \nonumber \\
	\mt{P} \int_{x_2}^\infty \mt{d} x_1 \frac{x_1 - x_2}{x_1^3 x_2^2 (x_1^2 - x_2^2)} &=& - \frac{1}{2 x_2^5} + \frac{\ln 2}{x_2^5}.
\end{eqnarray}
Hence, we obtain
\begin{equation}
	n_2^{\mt{RT}} (\omega_2) = \frac{c \hbar \alpha_2}{\pi t_{nn}} \left( \frac{t_{nn}}{\hbar \omega_2}\right)^5 \left( 2\ln 2 + \ln \frac{t_{nn}}{\hbar \omega_2} - \frac{1}{2} \right).
\end{equation}
We note that this quantity is positive if $\hbar \omega_2 > 0.41 t_{nn}$, and otherwise negative. 

To calculate the contribution from self-coupling, we need to evaluate
\begin{equation}
	n_2^{\mt{SC}} (x_2) = - \frac{4 c \hbar \alpha_2}{\pi t_{nn}} \left( \frac{1}{x_2}\right)^4 \mt{P} \int_0^\infty \frac{\mt{d} x_1}{x_1^2 - x_2^2} f_{vc} (t_{nn} x_1).
\end{equation}
After approximating $f_{vc} (x) \simeq \theta(x)$, it is easy to verify that the above integral vanishes.

Next, we evaluate the contribution from the quadratic AC Stark shift, which requires the evaluation of the following integral.
\begin{eqnarray}
	n_2^{\mt{QSS}} (x_2) = - \frac{c \hbar \alpha_2}{\pi t_{nn} x_2^2} \mt{P} \int_0^\infty &&  \frac{\mt{d} x_1}{x_1 (x_1 - x_2) (x_1^2 - x_2^2)} \nonumber \\
	&& \times f_{vc} (t_{nn} x_1).
\end{eqnarray}
We first approximate $f_{vc} (x) \simeq \theta(x)$. Next, we perform the Cauchy principal value integral with respect to $x_1$ in the two separate intervals $[x_0, x_2 - \epsilon]$ and $[x_2 + \epsilon, \infty)$. Adding the results together, discarding the singularity occuring at $x_1 = 0$, and also sending $x_0 \rightarrow 0$ and then $\epsilon \rightarrow 0$ produces a converging analytic result, except for one term involving $1/(x_2^2 \epsilon)$. The origin of this term is the divergence of the quadratic AC Stark shift near $x_2 \approx x_1$ in \refeq{eq:AC-stark}, which is unphysical since the form we used is valid only when the detuning is large. With a more physically accurate formula involving the natural line broadening of the upper state, it is well known that the AC Stark shift for the case $x_2 = x_1$ is zero. Moreover, the contribution from $x_1 = x_2^-$ is exactly canceled out by the contribution from $x_1 = x_2^+$. Hence, in the Cauchy principal value integral we discard this diverging term. The obtained result is
\begin{equation}
	n_2^{\mt{QSS}} (\omega_2) = \frac{c \hbar \alpha_2}{\pi t_{nn}} \left( \frac{t_{nn}}{\hbar \omega_2}\right)^5 \left( \ln \left(\frac{t_{nn}}{\hbar \omega_2}\right) + \frac{1}{2}\right).
\end{equation}

Let us define a lumped parameter
\begin{equation}
	\beta_2 \equiv \frac{c \hbar \alpha_2}{\pi t_{nn}} = 1.812 \times 10^{-18} ~(\mt{m}^2/\mt{W}).
\end{equation}
We can represent the various Kerr nonlinear coefficients using the following convention:
\begin{equation}
	n_2 (\omega_2) = \beta_2 G \left(\frac{\hbar \omega_2}{t_{nn}} \right). \label{eq:Kerr-coefficient}
\end{equation}
\begin{table}%[H] add [H] placement to break table across pages
	\caption{Summary of components for the Kerr nonlinear coefficient function $G$ in \refeq{eq:Kerr-coefficient} for various two-photon contributing mechanisms.\label{tab:Kerr}}
	\begin{ruledtabular}
		\begin{tabular}{cc}
			Contribution & $G(x)$ \\*[2mm]
			\hline
			\\*[-3mm]
			Two photon absorption & $- 4 \cfrac{\ln x}{x^5} $\\*[3mm]
			Raman transition & $\cfrac{1}{x^5} \left(2 \ln 2 - \ln x - \cfrac{1}{2} \right)$ \\*[3mm]
			Self-coupling & $0$\\*[3mm]
			Quadratic AC Stark shift & $ \cfrac{1}{x^5} \left(\cfrac{1}{2} - \ln x\right)$
			% Lines of table here ending with \\
		\end{tabular}
	\end{ruledtabular}
\end{table}

\begin{figure}[!tb]
	\centering
	\includegraphics[width=0.5\textwidth]{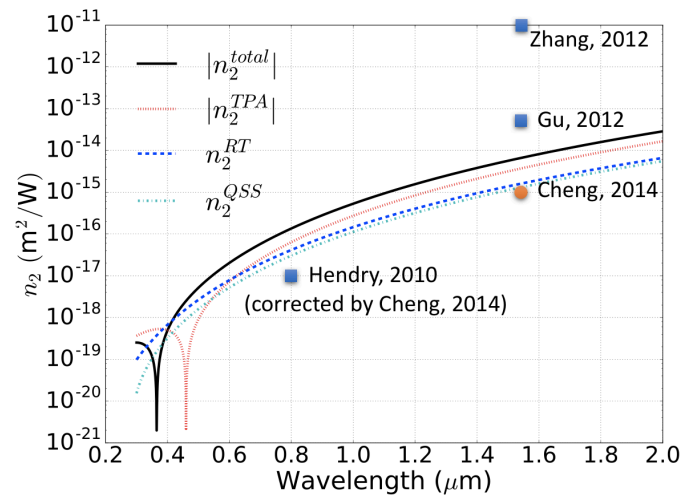}
	\caption{Kerr nonlinear coefficients from various mechanisms. Also the comparison with the previously reported values are shown. Blue squares: experiments, orange circle: theory. }
	\label{fig:comparison_n2}
\end{figure}

\begin{figure}[!tb]
	\centering
	\includegraphics[width=0.49\textwidth]{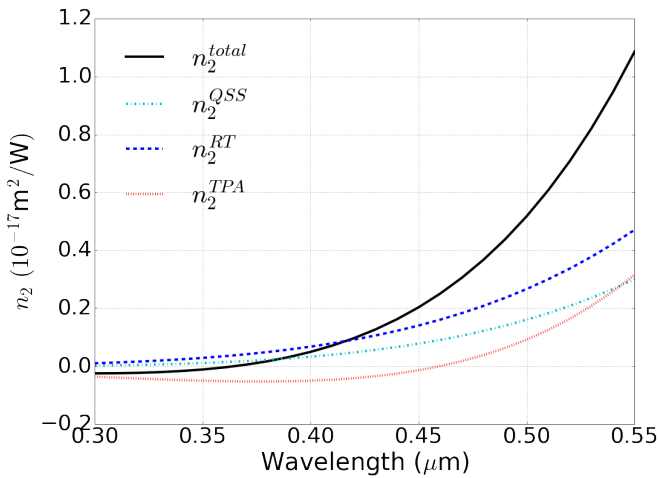}
	\caption{Kerr nonlinear coefficients at shorter wavelengths.}
	\label{fig:n2_shorter}
\end{figure}

TABLE \ref{tab:Kerr} summarizes the results. When added together, the total Kerr nonlinear coefficient becomes
\begin{equation}
	n_2^{\mt{total}} (\omega_2) = \beta_2 \left(\frac{t_{nn}}{\hbar \omega_2}\right)^5 \left[ 2 \ln 2  + 6  \ln \left(\frac{t_{nn}}{\hbar \omega_2}\right) \right]. \label{eq:final-result}
\end{equation}
We note that $n_2^{\mt{total}}$ changes sign at $\hbar \omega_2 = 1.26 t_{nn}$ (equivalently at 365 nm). If $\hbar \omega_2 > 1.26 t_{nn}$, the Kerr coefficient is positive and, otherwise, negative.  FIG \ref{fig:comparison_n2} shows the comparison of the magnitudes of the contributing mechanisms to the Kerr nonlinear coefficient. At wavelengths longer than 0.6 $\mu$m, the contribution from the two photon absorption is the largest. However, contributions from the Raman transitions and the quadratic AC stark effect are not small. It is clearly shown that the Kerr nonlinear coefficient is not singly determined by the two-photon absorption. A comprehensive calculation is thus necessary. FIG \ref{fig:n2_shorter} shows the close-up viewgraph for shorter wavelengths where the sign change occurs. At 365 nm, the total Kerr nonlinear coefficient vanishes. The contribution from the two-photon absorption is negative up to 455 nm while other contributions are positive.

\section{Discussion}

We first discuss the valid region of our calculations. Recall that we have assumed the linearized dispersion around the Dirac cones, which is explicitly used in deriving the analytic velocity matrix element in \refeq{eq:velocity-matrix}. The Kramers-Kronig relation in \refeq{eq:KK-nonlinear} integrates $\omega_1 \in [0,\infty)$. FIG \ref{fig:band1} in the appendix clearly shows that the band transition energy $\hbar \omega_1$ larger than 4 eV (equivalently, wavelengths shorter than 310 nm) deviates from the linear dispersion. Hence, one must resort to numerical integration of the Cauchy principal integral, in order to address the wavelengths below 310 nm. Additionally, one clear consequence of our linearized assumption used in the Cauchy principal value integral is that the Kerr nonlinear coefficient from the self-coupling became zero. If we take the same approach to calculate the linear refractive index $n_0$ through the Kramers-Kronig relation \refeq{eq:KK}, utilizing the linear absorption in \refeq{eq:linear-absorption}, it is straightforward to verify that we obtain $n_0 = 1$. However, a known experimental result exhibited that the refractive index of the graphene approaches that of the graphite $n_0 \approx 2.6$ \cite{bao2012graphene}. Hence, it is expected that our result may be different from the true values by a factor of unity magnitude. To compensate for this error, one may attempt to introduce a distortion function inside the Cauchy principal value integration for the high frequency components and artificially impose the consistency with the experimental result. However, this approach is somewhat discouraged since it is difficult to make an explicit formula for such a distortion function and carry out consistently all the other Cauchy principal value integrations for higher order mechanisms. Furthermore, our formalism did not include the participation of other higher or lower energy bands that may have small effects. Therefore, it should be clearly understood that our analytical final result in \refeq{eq:final-result} provides rather an order of magnitude estimate. 

We compare our theoretical result with previously reported results. We carefully selected the results that are only relevant to the Kerr nonlinear coefficient $n_2$, particularly excluding the results based on the third-harmonic generations, being a different mechanism from the Kerr effect. FIG \ref{fig:comparison_n2} shows the comparison. Firstly, for experimental results, Hendry \it et al. \rm originally reported $n_2 \simeq 1 \times 10^{-7}$ (esu) at 800 nm (equivalently $1.6 \times 10^{-14}$ (m$^2$/W)) using four-wave mixing technique \cite{hendry2010coherent}. However, Cheng \it et al. \rm \cite{cheng2014third} pointed out that the calibration of this result was not correct and recalculated based on the correct calibration, bringing down Hendry's result to $1 \times 10^{-17} $ (m$^2$/W), which agrees with our result in an order of magnitude estimate. Zhang \it et al. \rm reported $n_2 \simeq 10^{-11}$ (m$^2$/W) using Z-scan method at 1.55 $\mu$m \cite{zhang2012z}, which is three orders of magnitude larger than our theory. This experiment was performed on a graphene sitting on a quartz substrate, and it was hard to distinguish the pure graphene response from the combined response of the graphene, the interface, and the substrate compound. On the other hand, the result of Gu \textit{et al.} in \cite{gu2012regenerative} was obtained from a hybrid waveguide consisting of silicon and a mono-layer graphene. Utilizing the four-wave mixing spectra, Gu \textit{et al.} measured the combined Kerr nonlinear coefficient of the entire waveguide as $n_2 = 4.8 \times 10^{-17}$ (m$^2$/W) at 1.56 $\mu$m. Utilizing their calculation method based on the field filling-factors, one can reverse-estimate the sole contribution of the mono-layer graphene. The resulting estimate of a mono-layer graphene is  $n_2 = 6.2 \times 10^{-14}$ (m$^{2}$/W), which agrees with our theoretical result within an order of magnitude. For theoretical results, Cheng \it et al. \rm calculated $n_2 \sim 10^{-15}$ (m$^2$/W) at 1.56 $\mu$m as an order of magnitude estimate, which agrees well with our result. We note that Cheng \it et al. \rm considered only the two-photon absorption mechanism omitting the Raman transition and the quadratic AC Stark shift. 

\begin{figure}[!tb]
	\centering
	\includegraphics[width=0.5\textwidth]{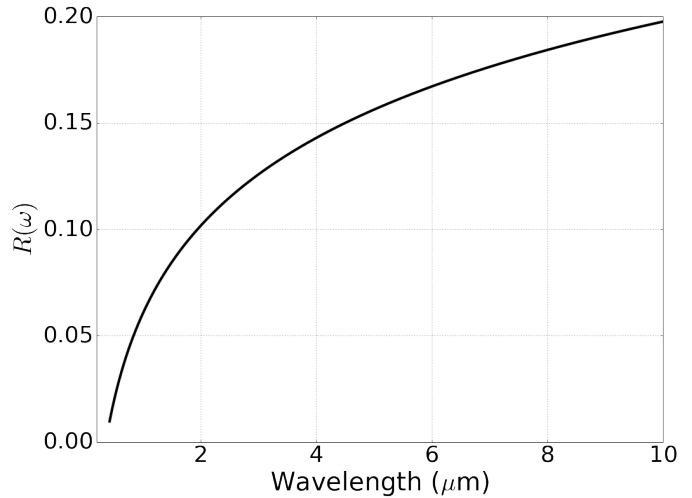}
	\caption{Plot of $R(\omega) = \chi'^{(3)} (\omega)/\chi''^{(3)}(\omega)$.}
	\label{fig:R}
\end{figure}

Another interesting quantity is the ratio between the real third-order susceptibility $\chi'^{(3)} (\omega)$ and the imaginary third-order susceptibility $\chi''^{(3)} (\omega)$. This ratio is quite important since $\chi'^{(3)} (\omega)$ is related to the coherent quantum evolution whereas $\chi''^{(3)} (\omega)$ is related to the incoherent decay. The relation among $\chi'^{(3)}$,$\chi''^{(3)}$, $\Delta \alpha_r$, and $n_2$ are \cite{sheik2001osa}
\begin{equation}
	n_2 = \frac{3}{4 \varepsilon_0 n_0^2 c} \chi'^{(3)}, \quad \Delta \alpha = \frac{3 \omega}{2 \varepsilon_0 n_0^2 c^2} \chi''^{(3)}.
\end{equation}
Then, we obtain
\begin{equation}
	\frac{\chi'^{(3)} (\omega)}{\chi''^{(3)} (\omega)} = \frac{2 \omega n_2 (\omega)}{c \Delta \alpha_r (\omega)}.
\end{equation}
For $\Delta \alpha_r (\omega)$, we only take the degenerate two-photon absorption as other mechanisms are irrelevant (see TABLE \ref{tab:absorption}). Then, we obtain
\begin{equation}
	R(\omega) \equiv \frac{\chi'^{(3)} (\omega)}{\chi''^{(3)} (\omega)} = \frac{\omega \beta_2}{16 c \alpha_2} \left( \frac{t_{nn}}{\hbar \omega}\right) \left[ \ln 2 + 3 \ln \left( \frac{t_{nn}}{\hbar \omega}\right)  \right].
\end{equation}
FIG \ref{fig:R} shows the plot of $R(\omega)$. At shorter wavelengths, the incoherent two-photon absorption clearly dominates whereas at longer wavelengths the coherent transition picks up and become significant. 

\section{Conclusion}

We reported a comprehensive theoretical analysis on the Kerr nonlinear coefficients of graphene at optical wavelengths. Our method is based on the Volkov-Keldysh-type dressed state treatment of the lattice wavefunctions, which allowed the first order perturbation to describe all relevant mechanisms that contribute to the Kerr nonlinear coefficient, namely, the two-photon absorption, the Raman two-photon transition, the self-coupling, and the quadratic AC Stark shift effect. We obtained a neat analytical formula for $n_2$. If desired, one can obtain more accurate result by numerical calculation using the full-band model, based on our formalism. Nevertheless, our analytical solution provides an order-of-magnitude estimate of the Kerr nonlinear coefficient. When compared with previous results, our theory predicts around the average order-of-magnitudes of the reported results, partly due to the widely disagreeing experimental results caused by the complexity of the measurement setup. We also analyzed the ratio between the coherent and the incoherent processes using the optical nonlinearity of the graphene, which suggests that long wavelengths in deep infrared or THz waves are much more preferred to perform cQED experiments using graphene.

\appendix
\section{Velocity matrix elements \label{sec:velocity-matrix}}

\begin{figure}[!tb]
	\centering
	\includegraphics[width=0.52 \textwidth]{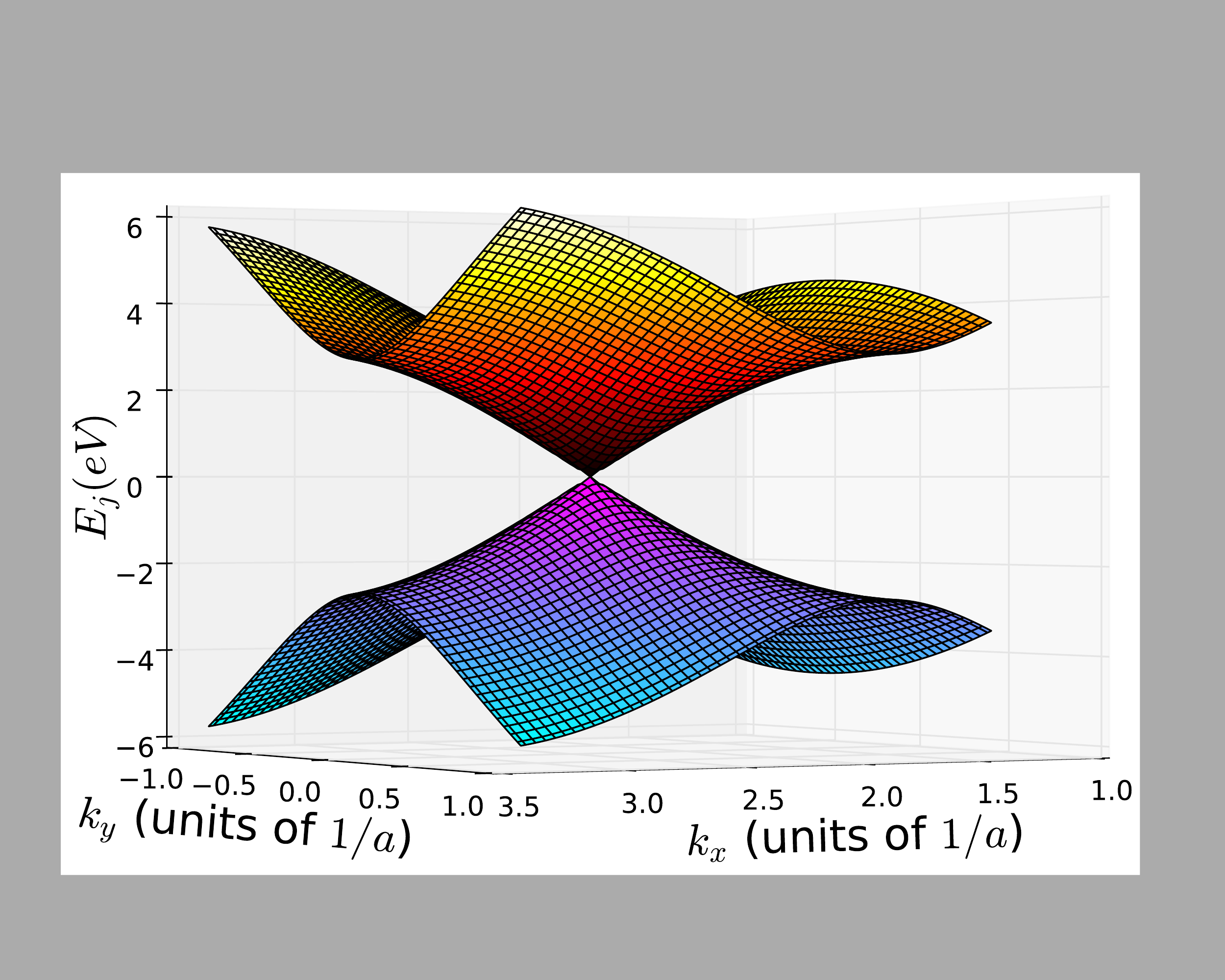}
	\caption{Energy dispersion around the Dirac cone at $\mathbf{K} = (4 \pi / (3 \sqrt{3} a), 0)$.} 
	\label{fig:band1}
\end{figure}

One uses the explicit formula for the Bloch function
\begin{eqnarray}
u_{j} (\kk, \rr) &&= \frac{1}{\sqrt{2N}} \sum_m \text{e}^{i \kk \cdot (\mathbf{R}_m - \rr)} \nonumber \\
&&  \times \left( s \frac{\gamma_{\kk}}{|\gamma_{\kk}|} \phi_{j} (\rr - \mathbf{R}_m) + \phi_{j} (\rr - \mathbf{R}_m - \mathbf{a}) \right), \label{eq:Bloch-function}
\end{eqnarray}
where $s = \pm 1$ if $j = c,v$ respectively, $\gamma_{\kk} = 1 + \text{e}^{i \kk \cdot \mathbf{a}_1} + \text{e}^{i \kk \cdot \mathbf{a}_2}$ is the structural factor with $\mathbf{a}_1 = \sqrt{3} a ((\sqrt{3}/2) \hat{\mathbf{x}} - (1/2) \hat{\mathbf{y}})$ and $\mathbf{a}_2 = \sqrt{3} a ((\sqrt{3}/2)  \hat{\mathbf{x}} + (1/2) \hat{\mathbf{y}})$ are the primitive lattice vectors, $a = 0.139$ nm is the distance between the nearest neighbor atomic sites, and $\phi_i(\rr)$ is the $|2p_z\rangle$ orbit wavefunction of a single atom centered at the origin. The energy eigenvalues are $E_{j} = s | \gamma_\kk | t_{nn}$ with the hopping energy between the nearest neighbor sites $t_{nn} = 2.7 $ eV \cite{neto2009electronic}. Figure \ref{fig:band1} shows the energy band structure of the graphene around one of the $\mathbf{K}$-points. Near the $\mathbf{K}$-point, the dispersion is circularly symmetric. Moreover, approximately up until the energy difference $E_c - E_i$ becomes 4 eV, the dispersion is quite linear. 

Threfore, we can linearize, around the $\mathbf{K}$-points, the structural factor to be $\gamma_{\kk} \approx (\hbar v_F / t_{nn}) (i q_x + q_y)$ where $(q_x, q_y) = \mathbf{q} = \mathbf{k} - \mathbf{K}$. Here $v_F$ is the Fermi velocity given by  
\begin{equation}
v_F = \frac{3 a t_{nn}}{2 \hbar}.
\end{equation}
The energy eigenvalues of the conduction and the valence bands become
\begin{equation}
E_c = + \hbar v_F |\mathbf{q}|, \quad E_v = - \hbar v_F | \mathbf{q}|. \label{eq:energy-eigenvalues}
\end{equation}
We first resolve the off-diagonal velocity matrix element $\mathbf{v}_{cv} = \langle \psi_c | \mathbf{v} | \psi_v \rangle$. Let us define the Berry phase
\begin{equation}
\bm{\xi}_{cv} \normalfont  \equiv \int_{R_0} \text{d}^2 r ~ u_c (\kk, \rr) \bm{\nabla}_{\kk} u_v (\kk, \rr).
\end{equation}
Using the Bloch function in \refeq{eq:Bloch-function}, one can show that 
\begin{equation}
\bm{\xi}_{cv} = - i \langle \psi_c (\kk) | \rr | \psi_v (\kk) \rangle. \label{eq:Berry-position}
\end{equation}
Using the basis $\{ |\Phi^A \rangle, | \Phi^B \rangle \}$ where $\langle \rr | \Phi^\alpha \rangle = (1/\sqrt{N}) \sum_m \text{e}^{i \kk \cdot (\mathbf{R}_m - \rr)} \phi_j (\rr - \mathbf{R}_m - \bm{\tau}^\alpha)$ with $\bm{\tau}^A = \mathbf{0}$ and $\bm{\tau}^B = \mathbf{a}$ is the vector from an atom in Bravais lattice $A$ to the nearest atom in Bravais lattice $B$, we can express the kets
\begin{equation}
|u_j (\kk) \rangle = \frac{1}{\sqrt{2}} \left( \begin{array}{c} s \gamma_\kk / |\gamma_\kk| \\ 1 \end{array} \right) = \frac{1}{\sqrt{2}} \left( \begin{array}{c} s\frac{i q_x + q_y }{\sqrt{q_x^2 + q_y^2}} \\ 1 \end{array}\right).
\end{equation}
where $s = \pm 1$ respectively for $j = c,v$. Using this representation, it is straightforward to calculate
\begin{equation}
\bm{\xi}_{cv} = \langle u_c (\kk) | \bm{\nabla}_\kk |u_v (\kk) \rangle = \frac{i \hat{\mathbf{z}} \times \mathbf{q}}{2 |\mathbf{q}|^2}, \label{eq:Berry-connector-graphene}
\end{equation}
where $\hat{\mathbf{z}}$ is the direction of light propagation, which is perpendicular to the graphene plane. The last step is to calculate using \refeq{eq:energy-eigenvalues}, \refeq{eq:Berry-position}, and \refeq{eq:Berry-connector-graphene}:
\begin{eqnarray}
\mathbf{v}_{cv} &=& \langle \psi_c (\kk) | \mathbf{v} | \psi_v (\kk) \rangle = - \frac{i}{\hbar} \langle \psi_c (\kk) | [\rr,  H_0 + H_{\text{int}}] | \psi_v (\kk) \rangle \nonumber \\
&=& \frac{1}{\hbar} (E_v - E_c) \bm{\xi}_{cv} = - i v_F \frac{\hat{\mathbf{z}} \times \mathbf{q}}{|\mathbf{q}|}.
\end{eqnarray}
In addition, it is straightforward to verify that $\mathbf{v}_{vc} = \mathbf{v}_{cv}^*$.

For the diagonal terms, we utilize the following group velocity of the Bloch wave:
\begin{equation}
\mathbf{v}_{ii} = \frac{1}{\hbar} \nabla_{\kk} E_i (\kk),
\end{equation}
where $i = c,v$. This immediately produces the result in \refeq{eq:velocity-matrix}.

\begin{acknowledgments}
	R. Hamerly is supported by a seed grant from the Precourt Institute for Energy at Stanford.
	
	Sandia National Laboratories is a multi-program laboratory managed and operated by Sandia Corporation, a wholly owned subsidiary of Lockheed Martin Corporation, for the U.S. Department of Energy’s National Nuclear Security Administration under contract DE-AC04-94AL85000.
\end{acknowledgments}

\bibliography{Gref2}

\end{document}